\documentclass[aps,prd,
nofootinbib,
preprintnumbers,
twocolumn,
showpacs, 
floatfix,
superscriptaddress
]{revtex4-1}
\usepackage{amsmath}
\usepackage{amssymb}
\usepackage{epsfig}
\usepackage{graphicx}
\usepackage{multirow}
\usepackage{color}
\usepackage[normal]{subfigure}
\usepackage{rotating}
\usepackage{hyperref}

\begin{document}

\thispagestyle{empty}

\preprint{ULB-TH/13-01}

\title{Constraints on primordial black holes 
as dark matter candidates from capture by neutron stars}


\author{Fabio Capela}
\email{fregocap@ulb.ac.be}
\affiliation{Service de Physique Th\'{e}orique, 
Universit\'{e} Libre de Bruxelles (ULB),\\CP225 Boulevard du 
Triomphe, B-1050 Bruxelles, Belgium}

\author{Maxim Pshirkov}
\email{pshirkov@prao.ru}
\affiliation{Sternberg Astronomical Institute, 
Lomonosov Moscow State University, 
Universitetsky prospekt 13, 119992, Moscow, Russia}

\affiliation{Pushchino Radio Astronomy Observatory, 
Astro Space Center, Lebedev Physical Institute  Russian Academy of Sciences,  
142290 Pushchino, Russia}

\affiliation{Institute for Nuclear Research of the Russian Academy of Sciences, 117312, 
Moscow, Russia}

\author{Peter Tinyakov}
\email{petr.tiniakov@ulb.ac.be}
\affiliation{Service de Physique Th\'{e}orique, 
Universit\'{e} Libre de Bruxelles (ULB),\\CP225 Boulevard du 
Triomphe, B-1050 Bruxelles, Belgium}


\begin{abstract} We investigate constraints on primordial black holes (PBHs) as dark
  matter candidates that arise from their capture by neutron stars (NSs).  If
  a PBH is captured by a NS, the star is accreted onto the PBH and gets
  destroyed in a very short time. Thus, mere observations of NSs put limits on
  the abundance of PBHs. High DM densities and low velocities are required to
  constrain the fraction of PBHs in DM. Such conditions may be realized in the
  cores of globular clusters if the latter are of a primordial origin.
  Assuming that cores of globular clusters possess the DM densities exceeding
  several hundred GeV/cm$^3$ would imply that PBHs are excluded as comprising
  all of the dark matter in the mass range $3\times 10^{18} \text{g} \lesssim
  m_\text{BH}\lesssim 10^{24} \text{g}$. At the DM density of $2\times
  10^3$~GeV/cm$^3$ that has been found in simulations in the corresponding models,
  less than 5\% of the DM may consist of PBH for these PBH masses.  
\end{abstract}



\maketitle

\section{Introduction}
\label{sec:introduction}

The existence of the dark matter (DM) has been established so far only
through its gravitational interaction. Consequently, little is known
about the DM nature apart from the fact that it is non-baryonic,
non-relativistic, weakly interacting and constitutes about $26.8\%$ of
the total energy budget of the Universe (for a recent review see,
e.g., \cite{Ade:2013lta,Bertone:2004pz,Bergstrom:2012fi}). 

Various candidates for the DM have been considered in the
literature. In the context of particle physics they are associated
with new stable particles beyond the Standard Model, a 
popular example being the so-called Weakly Interacting Massive
Particles (WIMPs). However, candidates that do not require new
stable particles also exist and are still viable. An attractive
candidate of this type is primordial black holes (PBHs)
\cite{Frampton:2010sw,Hawkins:2011qz}. This is the possibility we
consider in this paper. 

In the early universe, some primordial density fluctuations could have
collapsed producing a certain amount of black holes. These PBHs
possess properties that make them  viable DM candidates: they are
nonrelativistic and have a microscopic size of the order $r\sim
10^{-8} \text{cm}\; (m_\text{BH}/10^{20} \text{g})$, which makes them
effectively collisionless. The initial mass function of PBHs depends on their 
production mechanism in the early universe and is,
essentially, arbitrary. 
 
There exist a number of observational constraints on the fraction of PBHs
in the total amount of DM. First,
PBHs with masses $m_{\text{BH}}\leq 5\times 10^{14} \text{g}$
evaporate due to Hawking radiation \cite{Hawking:1974rv} in a time
shorter than the age of the Universe and cannot survive until today.
At slightly larger masses, even though the PBH lifetime is long
enough, the Hawking evaporation  still poses a  problem: the PBHs emit
$\gamma$-rays with energies around $100 \text{MeV}$ \cite{Page:1976wx}
in the amount that contradicts the data on the extragalactic gamma-ray
background. For instance, the Energetic Gamma Ray Experiment Telescope
 \cite{Sreekumar:1997un} 
has put an upper
limit on the cosmological density $\Omega_{\text{PBH}}\leq10^{-9}$ for
$m_\text{BH}=10^{15} \text{g}$ \cite{Carr:2009jm}. From such
observations, one can infer that PBHs with masses $m_{\text{BH}}\leq
10^{16} \text{g}$ cannot constitute more than $1\%$ of the DM. In the mass
range between $\sim 10^{18}$g and $\sim 10^{20}$g the PBH fraction is
constrained to less than 10\% by the femto-lensing of the gamma-ray 
bursts~\cite{Barnacka:2012bm}. 
More massive PBHs were constrained by EROS microlensing survey 
and the MACHO collaboration, which
set an upper limit of $3\%$ on the fraction of PBHs in the Galactic halo 
in the mass range $10^{26} \text{g}<m_{\text{BH}}<10^{30}\text{g}$
\cite{Tisserand:2006zx,1998ApJ...499L...9A}.
These constraints may be improved in the 
future~\cite{Griest:2011av,Cieplak:2012dp}. 
At even larger masses $10^{33} \text{g}<m_{\text{BH}}<10^{40}\text{g}$,
the three-year Wilkinson Microwave Anisotropy Probe (WMAP3) data 
and the COBE Far Infrared Absolute Spectrophotometer (FIRAS)
 data have been used to put
limits on the abundance of PBHs \cite{Ricotti:2007au}. These
constraints are summarized in Fig.~\ref{fig:constraints}. They leave
open the windows of masses 
$\mbox{(a few)}\times 10^{16} \text{g}<m_{\text{BH}}<10^{18}\text{g}$ and 
$10^{20} \text{g}<m_{\text{BH}}<10^{26}\text{g}$. 

In order to put constraints on PBHs in the remaining allowed mass range, in
Ref.~\cite{Capela:2012jz} we have considered the capture of PBHs by a star
during  star formation process and their further inheritance by the star's
compact remnant, the neutron star (NS) or the white dwarf (WD). The
presence of even a single PBH of a corresponding mass inside the
remnant (NS or WD) leads to a rapid destruction of the latter by
the accretion of the star matter onto the PBH
\cite{Kouvaris:2010jy,Kouvaris:2011fi,Kouvaris:2011gb,Kouvaris:2010vv,Kouvaris:2012dz}.
Thus, mere observations of NSs
and WDs in a DM-rich environment, such as could be present in the centers of globular
clusters, impose constraints on the fraction of PBHs in the DM and
exclude PBHs as the only DM candidate in the range of masses
$10^{16} \text{g}<m_{\text{BH}}<3\times 10^{22}\text{g}$. Still, a
range of PBHs masses from $3\times 10^{22} \text{g}$ to $10^{26}
\text{g}$ remains unconstrained.

In this paper we derive constraints that arise from the direct
capture of PBHs by NSs. The origin of the
constraints is the same as in Ref.~\cite{Capela:2012jz}: even a
single PBH captured by a compact star rapidly destroys the latter, so
the existing observations of the NSs and WDs require that the
probability of capture is much less than one. This implies constraints 
on the PBH abundance at the location of the compact star and may be translated
into constraints on the fraction of PBHs in the total amount of DM. 

Similarly to the constraints derived from the PBH capture during star
formation in Ref.~\cite{Capela:2012jz}, the constraints that follow from the
direct capture require a high DM density and low velocity dispersion, as may
be present in the cores of metal-poor globular clusters if the latter are of a primordial
origin. Within
the same assumptions, the main one being that
the cores of the globular clusters contain the DM density exceeding several
hundred GeV/cm$^3$ as is expected from numerical simulations (see Sect.~\ref{sec:constr} for a detailed discussion), we find that
the arguments based on the capture of PBHs by the NSs allow one to extend the
constraints of Ref.~\cite{Capela:2012jz} to higher PBH masses and exclude PBHs
as comprising 100\% of the DM up to $m_{\text{BH}} \lesssim \mbox{(a
  few)}\times 10^{24}$~g, leaving open only a small window of less than two
orders of magnitude.  Also, the constraints on the fraction $\Omega_{\rm
  PBH}/\Omega_{\rm DM}$ of PBHs in the total amount of DM at large PBH masses
become tighter as compared to Ref.~\cite{Capela:2012jz}. The final situation
is summarized in Fig.~\ref{fig:constraints}.

The rest of this paper is organized as follows.  In Sect.~\ref{sec:capture} we
discuss the capture of PBHs by compact stars. In
Sect.~\ref{sec:constr} we derive the constraints on the fraction of PBHs in
the DM from the capture in NSs. In Sect.~\ref{sec:concl} we summarize the
results and present our conclusions. Appendix~\ref{appendix} contains the calculation
of the energy loss by a BH passing through a neutron star.  Throughout the
paper, we use the units $\hbar = c=1$

\section{Capture of black holes by compact stars}
\label{sec:capture}

\subsection{Energy Loss}
\label{sec:energy-loss}

A PBH is captured if, during its passage through a star, it loses its
initial energy and becomes gravitationally bound. From this moment
every subsequent PBH orbit will again pass through the star, so that
finally the PBH will lose enough energy and will remain inside the
star all the time. 
Therefore, the criterion of
capture of a PBH is $E_\text{loss}> m_{\rm BH} v_0^2/2$ with
$E_\text{loss}$ being the energy loss during the collision and $v_0$ the
PBH asymptotic velocity. 
Two mechanisms of energy loss are operating during the collision:
deceleration of the PBH due to the accretion of star's material and
the so-called dynamical friction
\cite{1949RvMP...21..383C,1987gady.book.....B}. 
In the relevant range of PBH masses the accretion is less efficient 
compared to the
dynamical friction in the case of WDs, while the two mechanisms are
competitive in the case of NSs. 

As a PBH passes through the star, it transfers momentum and energy to the
surrounding matter. The result, called the dynamical friction, is a net force
that is opposite to the direction of motion of the PBH. As long as the PBH
velocity $v$ during the collision is larger than the velocity of the particles
constituting the compact object (which is a good approximation in the case of 
compact
stars), one may take the dynamical friction force to be
 \begin{equation}
 \textbf{f}_\text{dyn}=-4\pi G^2 m_\text{BH}^2 \rho \ln{\Lambda} 
\frac{ \textbf{v}}{v^3},
 \label{eq:dyn}
 \end{equation}
where $\rho$ is the density of the star matter and the factor $\ln(\Lambda)$ is
the so-called Coulomb logarithm
\cite{1949RvMP...21..383C,1987gady.book.....B} whose value is $\sim
30$ in the case of ordinary stars. Assuming a uniform
flux of incoming PBHs across the star, the average energy loss can be
written as follows, 
\begin{equation}
E_{\rm loss} = {4G^2 m_\text{BH}^2 M  \over R^2} \left\langle
{\ln{\Lambda} \over v^2} \right\rangle,
\label{eq:average-loss}
\end{equation}
where $M$ and $R$ are the mass and the radius of the star,
respectively, and $\langle...\rangle$ denotes the density-weighted
average over the star volume: 
\begin{equation}
\langle f(r) \rangle \equiv {1\over M} \int_0^R 4\pi r^2 dr \,\rho(r) f(r).
\label{eq:vol-average}
\end{equation}
When deriving eq.~(\ref{eq:vol-average}) we have transformed the integral 
along the PBH trajectory inside the star and the integral over the
orthogonal plane which comes from the averaging into a single 
integral over the star volume. We also 
accounted for the dependence of the velocity $v$ on the distance $r$ from the star
center, and allowed for an analogous dependence of the Coulomb
logarithm $\ln{\Lambda}$, as will be important in what follows.  

Taking into account that the PBHs velocity during the collision is of
order $v=v_\text{esc}=\sqrt{2GM/R}\gg v_0$, and assuming that
$\ln{\Lambda}$ is $r$-independent, the energy loss is parametrically given by
$E_\text{loss}\propto G m_\text{BH}^2/R$.  Since $E_\text{loss}$ is inversely
proportional to the radius of the star, NSs induce a much larger energy
loss during one collision compared to WDs. Thus, we will only
consider the case of NSs from now on.

Several complications arise in the calculation of $E_\text{loss}$ in the case
of NS. First, the accretion of the nuclear matter onto the PBH contributes
significantly into slowing it down. As far as the capture criterion is
concerned, the effect of the accretion can be incorporated into
eq.~(\ref{eq:average-loss}) by adding an extra contribution to the Coulomb
logarithm $\ln\Lambda \to \ln\Lambda(r) = \ln\Lambda + c(r) v^4$, where $c(r)$
is an $r$-dependent coefficient whose precise value is given in the
Appendix~\ref{appendix}.

Second, the core of a neutron star is comprised of the
degenerate neutron gas, so the question arises to which extent 
eq.~(\ref{eq:dyn}) is still applicable. 
Here we note that by the time the falling PBH reaches the core of NS it picks a
relativistic velocity $v\sim 0.6c$. This velocity is by a factor of a few
larger than the velocity of sound, so the nucleons can be considered as free
particles and the arguments leading to eq.~(\ref{eq:dyn}) apply. With this
velocity, the PBH 
can transfer to neutrons the momentum
of up to $\sim 1.8$~GeV, which is by a factor of a few larger than the Fermi
momentum of neutrons in the center of the star, and much larger than the Fermi
momentum away from the center. However, only neutrons with sufficiently small
impact parameters --- such that the momentum transfer is larger than their Fermi
momentum --- contribute to slowing the BH down. Thus, the Coulomb logarithm
gets cut at a much smaller distance which, moreover, depends on the 
local density of neutrons through their Fermi momentum.  

Both effects can be incorporated into eq.~(\ref{eq:average-loss}) through the
$r$-dependence of $\ln \Lambda$ and, finally, expressed in terms of 
the average value of 
$\langle \ln\Lambda /v^2\rangle$. We have calculated this quantity numerically
making use of a concrete NS density profile from Ref.~\cite{NSprofile} (see
Appendix~\ref{appendix} for details). We found
\begin{equation}
\left\langle {\ln \Lambda \over v^2} \right\rangle
= 14.7.
\label{eq:eff_lambda}
\end{equation}
As we argue in the Appendix~\ref{appendix}, this value depends weakly on the
NS mass and radius. Making use of eq.~(\ref{eq:eff_lambda}) one obtains
\begin{equation}
E_{\rm loss} /m_{\rm BH} = 6.3\times 10^{-12} \left({m_{\rm BH} \over
  10^{22}{\rm g}}\right),
\label{eq:Eloss-NS}
\end{equation}
where we have substituted $R=12$~km and $M=1.4\;M_\odot$ as typical NS
parameters. These values for the radius and the mass of the NS are assumed
throughout the rest of the paper except where the opposite is stated
explicitely.

It remains to be checked that, once the PBH becomes gravitationally
bound, multiple collisions bring the PBH inside the NS sufficiently fast. 
Assuming a radial orbit and denoting the apastron $r_{\rm max}$, the
half-period is 
\[
\Delta T = {\pi r_{\rm max}^{3/2}\over \sqrt{GM}}.
\]
The energy loss in half a period (that is, during a single collision with NS)
as a function of $r_{\rm max}$ is given by
eq.~(\ref{eq:average-loss}). Dividing the energy loss by the time and
expressing the energy in terms of $r_{\rm max}$ one obtains the differential
equation for the evolution of $r_{\rm max}$ as a function of time,
\begin{equation}
\dot \xi = - {1\over \tau} \sqrt{\xi}, 
\label{eq:xi-of-t}
\end{equation}
where $\xi=r_{\rm max}/R$ and 
\[
\tau = {\pi R^{5/2}\over 4 Gm_{\rm BH} \sqrt{GM}} 
\left\langle {\ln \Lambda \over v^2} \right\rangle^{-1} \simeq 
8\times 10^6 {\rm s} \left({m_{\rm BH}\over 10^{22}{\rm g} }\right)^{-1}.
\]
The corresponding energy loss time is 
\[
t_{\rm loss} \simeq 2 \tau \sqrt{\xi_0},
\]
where the initial value $\xi_0$ can be estimated by requiring that the  
initial PBH energy is
of the order of $E_{\rm loss}$. Assembling all the factors one has
\begin{equation}
t_{\rm loss} \simeq 4.1\times 10^4 {\rm yr} 
\left({m_{\rm BH}\over 10^{22}{\rm g} }\right)^{-3/2}.
\label{eq:tloss}
\end{equation}
Thus, PBHs heavier than $m_{\rm PBH}\gtrsim 2.5\times 10^{18}$~g end up 
inside the NS in a time shorter than $10^{10}$~yr.

\subsection{Capture Rate}
In order to calculate the capture rate, we assume that the PBHs follow
a Maxwellian distribution in velocities with the dispersion $\bar{v}$,
\begin{equation}
dn=n_\text{BH}\left(\frac{3}{2\pi \bar{v}^2}\right)^{3/2} 
\exp\left\{\frac{-3v^2}{2\bar{v}^2}\right\} d^3v,
\end{equation}
where $n_\text{BH}=\rho_\text{BH}/m_\text{BH}$, $\rho_\text{BH}$ being
the density of PBHs at the star location. It can be expressed in terms
of the local DM density $\rho_\text{DM}$ as follows, 
\begin{equation}
\rho_\text{BH} = {\Omega_\text{PBH}\over \Omega_\text{DM}}\rho_\text{DM} .
\label{eq:rhoBH}
\end{equation}
Following
\cite{Kouvaris:2007ay}, the capture rate takes the form
\begin{equation}
F = {\Omega_\text{PBH}\over \Omega_\text{DM}} F_0,
\label{eq:F=F_0}
\end{equation}
where 
\begin{equation}
F_0=\sqrt{6\pi}{\rho_\text{DM}\over m_\text{BH}} {R_gR\over \bar v (1-R_g/R)} 
\left(1-\exp\left(-\frac{3E_\text{loss}}{m_{\rm BH} \bar{v}^2}\right)\right)
\label{eq:capture}
\end{equation}
is the capture rate assuming PBHs comprise all of the DM, $R_g=2GM$ is the 
Schwarzschild radius of the NS and 
$E_{\text{loss}}$ is given by eq.~(\ref{eq:Eloss-NS}). 

Two different regimes are possible depending on the PBH mass. 
In the case when the energy loss is small,
$E_\text{loss}\ll m_{\rm BH} \bar{v}^2/3$, the exponential can be
expanded and one gets at the leading order
\begin{equation}
F_0=3\sqrt{6\pi}{\rho_\text{DM}\over m_\text{BH}} {R_gR \over \bar v^3(1-R_g/R)} 
{E_\text{loss} \over m_\text{BH}}.
\label{eq:Eloss-small}
\end{equation}
In view of eq.~(\ref{eq:Eloss-NS}) the capture rate is independent of 
$m_\text{BH}$ in this regime. 
In the opposite case $E_\text{loss}\gg m_{\rm BH} \bar{v}^2/3$ the
exponential in eq.~(\ref{eq:capture}) can be neglected and 
\begin{equation}
F_0=\sqrt{6\pi}{\rho_\text{DM}\over m_\text{BH}} {R_gR \over \bar v(1-R_g/R)},
\label{eq:Eloss-large}
\end{equation}
so that the capture rate decreases with increasing $m_\text{BH}$. In both
cases the capture rate is inversely proportional to some power of velocity and
is thus maximum for sites with high dark matter density $\rho_\text{DM}$ and
small velocity dispersion $\bar{v}$.

\section{Constraints}
\label{sec:constr}

As previously mentioned, if a NS captures a PBH, the accretion
of the NS material onto the PBH rapidly destroys the star. Therefore,
observations of NSs imply constraints on the capture rate of PBHs
which has to be such that the probability of the PBH capture is much
less than one. In view of eq.~(\ref{eq:F=F_0}) these constraints
translate into constraints on the fraction of PBHs in the dark matter, 
$\Omega_\text{PBH}/\Omega_\text{DM}$. 

Given a NS of age $t_{\rm NS}$, the probability of its survival is
$\exp(-t_{\rm NS}F)$ with $F$ given by eqs.~(\ref{eq:F=F_0}) and
(\ref{eq:capture}). Requiring that the survival probability is not small
leads to the constraint
\begin{equation}
\frac{\Omega_\text{PBH}}{\Omega_\text{DM}}
\leq  {1\over t_\text{NS}F_0}.
\label{eq:constraint}
\end{equation}
Depending on the environment where the NS is located, $F_0$ may vary
by many orders of magnitude. The most stringent constraints come
from sites where $F_0$ is high. Among such sites, globular
clusters (GCs) are the best candidates. 

GCs are compact, nearly spherical collections of
stars scattered over the Galactic halo. They have ages between 8 to
13.5 Gyr, and as such are the oldest substructures of our Galaxy.  GCs are
made of population II stars, WDs, NSs and black holes. A typical GC
has an average radius of 30 pc, a core radius of 1 pc and a baryonic
mass of $\mbox{(a few)} \times 10^5 M_\odot$ \cite{Dotter2010}.

The DM content of GCs is a matter of an ongoing debate.  The
distribution of metallicity in GCs is bimodal, indicating two
subpopulations formed by different mechanisms \cite{Brodie2006}. The
metal-rich GCs are considered to be formed during gas-rich mergers in
proto-galaxies~\cite{Fall1985,Ashman1992,Kravtsov2005,Muratov2010}. These GCs contain
very little DM, if any. Instead, as cosmological simulations show,  metal-poor GCs could  have been formed in low-mass dark matter halos at very high-redshift
$z\sim 10-15$~\cite{Peebles1984,Bromm2002,MS2005a,Moore2006,Boley2009,Griffen2010}.
Observations of GCs show no evidence of DM halos
\cite{Moore1996}. This is expected as the halos should have been
tidally stripped due to interactions with the Galaxy
\cite{MS2005b}. The DM content would, however, be preserved in the
cores of such GCs. In support of this picture, it has been found in
Refs.~\cite{MS2005b,MS2005a}, using high-resolution N-body
simulations, that many properties of simulated GCs with DM halos are
similar to those of observed GCs. In what follows we will 
focus on metal-poor GCs and assume that they have been formed in DM
halos and thus possess DM-rich cores. 
\begin{figure}
\begin{center}
\begin{picture}(250,160)
\put(0,0){\includegraphics[width=1.\columnwidth]{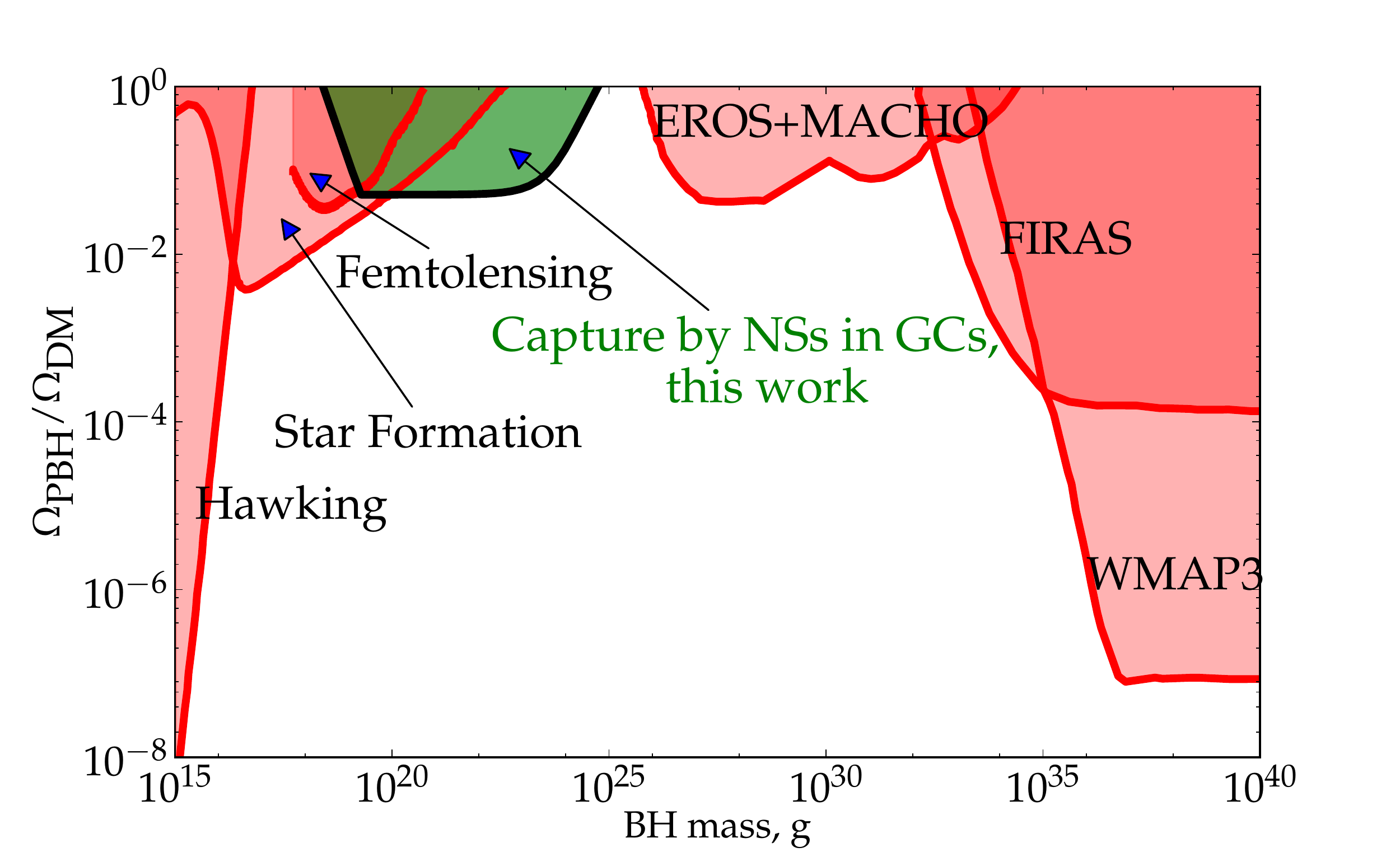}}
\end{picture}
\end{center}
\caption{\label{fig:constraints} Constraints on the fraction of PBHs in the
  total amount of DM from various observations as explained in
  Sect.~\ref{sec:introduction} (red shaded regions). The green shaded
  region shows constraints derived in this paper, which would follow from
  observations of NSs in the cores of globular clusters if one assumes
  the DM density of $2\times 10^3$~GeV/cm$^3$ as obtained in numerical
  simulations. 
}
\end{figure} 

In Ref.~\cite{Bertone:2007ae} the DM density close to the core of such GC
has been estimated to be of the order $\rho_\text{DM}\sim2\times10^3 \text{GeV}
\,\text{cm}^{-3}$.  This result was concluded to be rather
independent of the original halo mass and is in agreement with
N-body simulations \cite{MS2005a,MS2005b}. Therefore, we adopt this
value in our estimates.

The velocity dispersion is another important parameter. Since stars
are collisionless and therefore behave similarly to DM particles, this
parameter can be extracted from observations.  We adopt the value of
$\bar{v}=7~\text{km}~\text{s}^{-1}$. The velocity dispersion varies
noticeably from cluster to cluster. The list of measured velocities of known 
GCs can be found in 
Ref.\cite{1993ASPC...50..357P}; the adopted value is a median of this
distribution. 
Finally, we adopt the NS radius
$R_\text{NS}=12~ \text{km}$ and mass $M_\text{NS}=1.4~ M_\odot$ as stated
above, 
and
the life time $t_{\rm NS} = 10^{10} \text{yr}$ \cite{Tauris:2012cn}.

The constraints arising from observations of NSs in the core of a GC under
these assumptions, as well as previously existing constraints are summarized
in Fig.~\ref{fig:constraints}.  As one can see, the new constraints exclude
the PBHs as the unique DM component for masses lower than $m_\text{BH}\sim
\mbox{(a few)}\times10^{24} \text{g}$, thus extending by about two orders of
magnitude the constraints derived in Ref.~\cite{Capela:2012jz} to higher PBH
masses.

In qualitative terms, the shape of the exclusion region in
Fig.~\ref{fig:constraints} is easy to understand from
eqs.~(\ref{eq:Eloss-small}) and (\ref{eq:Eloss-large}). The horizontal part of
the curves is due to eq.~(\ref{eq:Eloss-small}) where the dependence on the
PBH mass cancels out (cf. eq.~(\ref{eq:Eloss-NS})). The inclined part on the
right results from eq.~(\ref{eq:Eloss-large}). The transition between the two
regimes is at the PBH mass such that $E_\text{loss}\sim m_{\rm BH}
\bar{v}^2/3$. The sharp cut at small masses occurs when the time needed for
multiple collisions to bring the PBH inside the NS exceeds the NS lifetime.

\begin{figure}
\begin{center}
\begin{picture}(250,160)
\put(0,0){\includegraphics[width=1.\columnwidth]{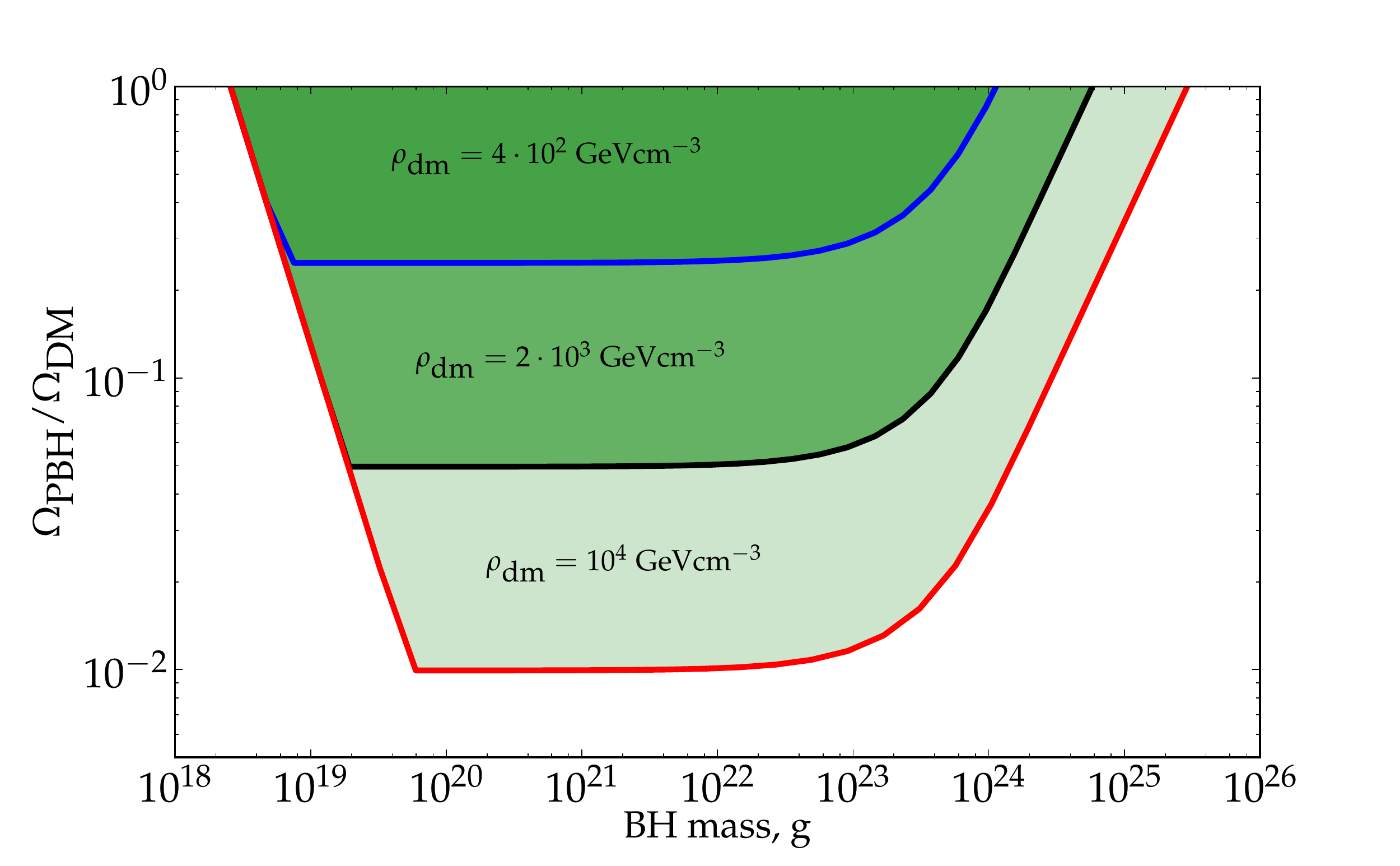}}
\end{picture}
\end{center}
\caption{\label{fig:constdens} The dependence of the constraints 
on the fraction $\Omega_\text{PBH}/\Omega_\text{DM}$ of PBHs in 
the total amount of DM on the assumed 
DM density in the core of a GC. Three cases are shown: 
$\rho_\text{DM}=4\times 10^{2}~\text{GeV}\text{cm}^{-3}$, 
$\rho_\text{DM}=2\times 10^{3}~\text{GeV}\text{cm}^{-3}$ (the same as in
Fig.~\ref{fig:constraints}), 
and $\rho_\text{DM}=10^{4}~\text{GeV}\text{cm}^{-3}$.} 
\end{figure}

Given the uncertain DM content of the GCs, in Fig.~\ref{fig:constdens}
we show the dependence of the constraints on the 
assumed DM density in the GC core. Apart from the cutoff at small masses, the
constraints scale trivially with the DM density. The dependence on the 
velocity dispersion is similar, but not identical (not shown in
Fig.~\ref{fig:constdens}): the horizontal part of the constraints scales like 
$1/\bar v^3$, while the inclined part at large masses scales like $1/\bar v$. 

\section{Conclusions}
\label{sec:concl}
We have studied the constraints on the fraction of PBHs in the total amount of
DM that arise from the requirement that PBHs be captured by NSs with
probability much less than one, since capture of even a single PBH leads to a
rapid accretion of the star matter onto the PBH and eventual star
destruction. High DM density in excess of several hundred GeV/cm$^3$ and low
velocity dispersion are required to obtain meaningful constraints. Such
conditions may be realized in the cores of metal-poor globular
clusters if they are formed in low-mass DM halos at very high-redshift $z \sim
10-15$.

If the metal-poor globular clusters are indeed of a primordial origin,
simulations predict that their cores have DM densities as high as $2\times
10^3$GeV/cm$^3$ \cite{Bertone:2007ae}. At this value, our constraints would
exclude PBH as the only DM candidate in the mass range $3\times 10^{18}
\text{g}\leq m_\text{BH}\leq 5\times10^{24}\text{g}$. Together with the
previously existing constraints, this would leave open only a small window of
masses around $10^{25}$~g where PBHs can still constitute all of the DM. Note,
however, that a viable PBH model would have to explain a very narrow PBH mass
distribution of the width of less than two orders of magnitude.

As one can see in Fig.~\ref{fig:constraints}, the 
constraints derived here are complementary to those of 
Ref.~\cite{Capela:2012jz}. The constrained region has been extended up
to
masses $\sim 5\times 10^{24}$g. While in
Ref.~\cite{Capela:2012jz} better constraints were achieved for masses
$10^{16} \text{g}\leq m_\text{BH}\leq 10^{20} \text{g}$, here
we obtain more competitive constraints for masses 
$m_\text{BH}\geq 10^{20} \text{g}$. It is also important to note
that different assumptions are required in the two cases: while the
constraints of Ref.~\cite{Capela:2012jz} are sensitive to the DM
distribution at the epoch of the GC formation, for the constraints
derived in this paper the present-epoch DM distribution in GCs is
relevant.

We did not present the constraints that come from observations of the
Galactic center, which is another relatively close region of high DM
density. If the DM density in the Galactic center is comparable to
that assumed above for the the cores of the GCs, no new constraints
arise from that region \cite{Abramowicz:2008df}. The reason is that 
the capture rate depends
strongly on the PBH velocity dispersion,
cf. eq.~(\ref{eq:Eloss-small}), which is by more than an order of
magnitude larger in the Galactic center than in the cores of GCs. It
has been suggested, however, that the DM density in the Galactic
center may be as high as $\rho_\text{DM}=10^{6} \text{GeV}~
\text{cm}^{-3}$ \cite{Bertone:2005xv}. If this were confirmed, the
constraints from the Galactic center would become competitive to the
ones presented here.

\acknowledgments

The authors are indebted to G. Rubtsov and A.~Gould for comments on
the first version of the manuscript.  M.P. acknowledges the hospitality of the
Service de Physique Th\'{e}orique of ULB where this work was
initiated. The work of F.C. and P.T. is supported in part by the IISN
and the Belgian Science Policy (IAP VII/37). The work of MP is supported 
by RFBR Grants No.~12-02-31776 mol\_a, No.~13-02-00184a, No.~13-02-01311a, 
No.~13-02-01293a, by the Grant of the President of Russian Federation MK-2138.2013.2 
and by the Dynasty Foundation.

\appendix*
\section{Calculation of the friction force}
\label{appendix}

When the BH moves through a neutron star, it experiences a friction force that
is the result of scattering and accretion of nucleons. In
Sect.~\ref{sec:capture} we have written this force in the form
(\ref{eq:average-loss}) analogous to the dynamical friction
\cite{1949RvMP...21..383C} with all the effects combined in the single factor
$\langle \ln \Lambda/v^2\rangle$. Here we calculate this factor.

To make the calculations manageable, we make a number of simplifying
assumptions: (i) We treat the motion of the BH through the NS in the Newtonian
approximation (that is, we neglect the general relativity effects), but do not
assume the BH to be non-relativistic. In fact, the BH in the center of the
star may attain velocities of up to about $0.6c$. (ii) Since the BH velocity
exceeds the sound speed, we treat the nucleons as free particles and account
only for their individual interactions with the BH. (iii) To determine which
neutrons of the degenerate matter of the NS are excited and absorb momentum we
use a simple criterion: we require that the momentum transferred to the
neutron in the gravitational collision with the BH exceeds its Fermi momentum
$k_F$.

In the BH reference frame, the scattering of a nucleon off the BH is
described by the following expression  \cite{landashitz} for 
the scattering angle
$\phi(b)$ as a function of the impact parameter $b$,
\begin{equation}
\phi(b) = -\pi + 2 \tilde b \int_0^{x_{\rm max}} 
{dx \over \sqrt{\gamma^2 -(1+\tilde b^2 x^2)(1-x)}},
\label{eq:phi}
\end{equation}
where $\gamma$ is the gamma factor of the nucleon, $\tilde
b=bv\gamma/R_g$ is the rescaled impact parameter, $R_g$ being the
gravitational radius of the BH, and $x_{\rm max}$ is the smallest
zero of the denominator in eq.~(\ref{eq:phi}). The variable $x$ is
the inverse distance between the nucleon and the BH in units of $R_g$,
so that in terms of the distance the integration range in
eq.~(\ref{eq:phi}) is from infinity to the point of the closest
approach. Eq.~(\ref{eq:phi}) includes all the GR effects. 

The scattering is impossible below some critical value of the impact
parameter $b_{\rm crit}$ which is determined by the set of equations
\begin{eqnarray}
\nonumber
\gamma^2 &=& U(x),\\
{\partial U \over \partial x} & = & 0,
\label{eq:b-crit}
\end{eqnarray}
where $U(x) = (1+\tilde b^2 x^2)(1-x)$.
For smaller values $b< b_{\rm crit}$ the nucleons get accreted onto
the BH. The value of $b_{\rm crit}$ depends only on the relative
asymptotic velocity of BH and nucleons $v$; at $v=0.6$ one has $b_{\rm
  crit}= 3.79 R_g$. 

Consider the case of scattering, $b> b_{\rm crit}$. In the
reference frame of the NS  the nucleons are initially at rest. After the
collision they acquire the momentum 
\begin{equation}
\Delta p = (mv\gamma^2 (-1+\cos\phi), m v \gamma \sin\phi, 0), 
\label{eq:p-transfer}
\end{equation}
$m$ being the neutron mass and we have assumed that the BH velocity 
is along the $x$-direction. 
The nucleons contribute to the friction force only up to some impact
parameter $b_{\rm max}$ which is determined by the equation 
\begin{eqnarray}
k_F^2 &\equiv& \left(3\pi^2{\rho\over m_n}\right)^{2/3} \nonumber \\
&=&m^2 v^2 \gamma^2 \left\{ (1-\cos\phi(b))^2\gamma^2 + \sin^2\phi(b)
\right\},
\label{eq:bmax}
\end{eqnarray}
where $\rho$ is the neutron density. Note that the resulting value of $b_{\rm
  max}$ depends on the nucleon density through the first equality of
eq.~(\ref{eq:bmax}).

After the collisions with many nucleons the $y$-component of the transferred
momentum averages away, while the $x$-component adds up and results in the
friction force acting on the BH. Including the effect of the accreted
nucleons, one can write this force as follows:
\begin{equation}
{d E \over dr} 
= 4 \pi \rho {G^2 m_{\rm BH}^2\over v^2} \ln \Lambda(r),
\label{eq:lambda_eff}
\end{equation}
where 
\begin{eqnarray}
\ln \Lambda(r) &=& v^4\gamma^2 {b_{\rm crit}^2\over R_g^2} \nonumber \\
&+& v^4\gamma^2 {2 \over R_g^2} \int_{b_{\rm crit}}^{b_{\rm max}}
b\, db (1-\cos\phi(b)) .
\label{eq:lambda_fin}
\end{eqnarray}
The first term in this expression is due to the accretion, while the
second to the scattering of nucleons. It is easy to check that in the 
non-relativistic limit and assuming non-degenerate matter (that is,
extending the integral to the size of the star), the second term
dominates and 
reduces to the standard expression for the Coulomb logarithm. 
Making use of eq.~(\ref{eq:lambda_fin}) the density-weighted average
in eq.~(\ref{eq:average-loss}) reads
\begin{eqnarray}
\left\langle {\ln \Lambda \over v^2} \right\rangle 
= {4\pi \over MR_g^2}&& \int_0^{R_{\rm NS}} r^2  dr \rho(r)v^2\gamma^2 
\Bigl\{ b_{\rm crit}^2 \nonumber\\
&+& 2  \int_{b_{\rm crit}}^{b_{\rm max}}
b\, db (1-\cos\phi(b)) \Bigr\}. 
\label{eq:<ln-Lambda>}
\end{eqnarray}
Here  $v$, $\gamma$, $b_{\rm crit}$ and $b_{\rm max}$ all depend on
$r$. Note that in view of eqs.~(\ref{eq:phi}), (\ref{eq:b-crit}) and 
(\ref{eq:bmax}) this equation is independent of the BH mass 
$m_{\rm BH}$. 

We have calculated this expression numerically. As an input we used
the tabulated NS density profile given in Ref.~\cite{NSprofile} which
corresponds to the NS of mass $1.8M_\odot$ and radius $13.5$~km. For a
given value of $r$ we have calculated $v$ and $\gamma$ in the
Newtonian approximation, determined the
critical impact parameter $b_{\rm crit}$ from eqs.~(\ref{eq:b-crit})
(the latter can be solved analytically), calculated the function
$\phi(b)$ from eq.~(\ref{eq:phi}) and the maximum impact parameter 
$b_{\rm max}$. 
We considered the NS matter to be degenerate down to densities $\rho =
10^{14}$~g/cm$^3$ which we took as the boundary of the NS
crust~\cite{NSprofile}. 
Finally, we have calculated the integral in 
eq.~(\ref{eq:<ln-Lambda>}) and found that it equals 14.7, which gives
eq.~(\ref{eq:eff_lambda}). 
The contributions of the accretion and dynamical friction (the
first and the second terms in eq.~(\ref{eq:<ln-Lambda>})) are roughly
equal. 

In conclusion, an important remark is in order. Although we have performed the
calculation for a concrete NS mass, the result depends very weakly on the
latter. We have checked this by rescaling the density profile of 
Ref.~\cite{NSprofile}
in such a way that the new NS mass and radius are $1.4M_\odot$ and 12~km,
respectively. Repeating the above calculations, we have found that the
average in eq.~(\ref{eq:<ln-Lambda>}) changes by less than 4\%. We neglect
this difference and use the value given in eq.~(\ref{eq:eff_lambda}) in our
estimates.

\bibliography{mainbibv1} 
\bibliographystyle{apsrev}

\end{document}